\documentclass[sn-nature]{sn-jnl}  
\usepackage{graphicx}%
\usepackage{multirow}%
\usepackage{amsmath,amssymb,amsfonts}%
\usepackage{amsthm}%
\usepackage{mathrsfs}%
\usepackage[title]{appendix}%
\usepackage{xcolor}%
\usepackage{textcomp}%
\usepackage{manyfoot}%
\usepackage{booktabs}%
\usepackage{algorithm}%
\usepackage{algorithmicx}%
\usepackage{algpseudocode}%
\usepackage{listings}%
\usepackage{algorithm, algpseudocode}

\usepackage{array}
\usepackage[thinlines]{easytable}
\usepackage{tabularx}

\usepackage{caption}
\usepackage{subcaption}
\usepackage{setspace}
\usepackage{comment}
\usepackage{siunitx}
\newsavebox{\measurebox}
\renewcommand{\deg}{^{\circ}}
\renewcommand{\deg}{^{\circ}}
\newcommand{\as}{^{\prime\prime}}

\def\as{\prime\prime}
\renewcommand{\deg}{^{\circ}}

\begin{document} 
\title[Article Title]{Spectroscopic Investigation of Nebular Gas (SING): Instrument Design, Assembly and Calibration}

\author*[1,2]{\fnm{Bharat} \sur{Chandra P.}}\email{bharat.chandra@iiap.res.in}

\author[1]{\fnm{Binukumar} \sur{G. Nair}}
\email{binukumar@iiap.res.in}
\author[1]{\fnm{Shubham Jankiram} \sur{Ghatul}}
\author[1]{\fnm{Shubhangi} \sur{Jain}}
\author[1]{\fnm{S.} \sur{Sriram}}
\author[1]{\fnm{Mahesh} \sur{Babu S.}}
\author[1]{\fnm{Rekhesh} \sur{Mohan}}
\author[1]{\fnm{Margarita} \sur{Safonova}}\email{margarita.safonova@iiap.res.in}
\author[1]{\fnm{Jayant} \sur{Murthy}}
\author[3]{\fnm{Mikhail} \sur{Sachkov}}

\affil*[1]{ \orgname{Indian Institute of Astrophysics}, \orgaddress{\city{Bangalore}, \country{India}}}
\affil[2]{\orgdiv{Department of Applied Optics and Photonics}, \orgname{University of Calcutta}, \orgaddress{\city{Kolkata}, \country{India}}}
\affil[3]{\orgname{Institute of Astronomy of the Russian Academy of Sciences}, \orgaddress{\city{Moscow}, \country{Russia}}}

\abstract{The Spectroscopic Investigation of Nebular Gas (SING) is a near-ultraviolet (NUV) low-resolution spectrograph payload designed to operate in the NUV range, 1400 \si{\angstrom } -- 2700 \si{\angstrom}, from a stable space platform. SING telescope has a primary aperture of 298 mm, feeding the light to the long-slit UV spectrograph. SING has a field of view (FOV) of 1°,  achieving a spatial resolution of 1.33 arc minute and spectral resolution of 3.7 Å ($R\sim600$) at the central wavelength. SING employs a micro-channel plate (MCP) with a CMOS readout-based photon-counting detector. The instrument is designed to observe diffuse sources such as nebulae, supernova remnants, and the interstellar medium (ISM) to understand their chemistry. 
SING was selected by the United Nations Office for Outer Space Affairs to be hosted on the  Chinese Space Station. The instrument will undergo qualification tests as per the launch requirements. In this paper, we describe the hardware design, optomechanical assembly, and calibration of the instrument.}


\maketitle

\section{Introduction}
\label{sect:intro} 

The ultraviolet (UV: 900 Å – 3000 Å) band of the electromagnetic spectrum has more absorption and emission line density than any other band, making it one of the most interesting regions of the astrophysical spectrum. Imaging observations excel at tracing the morphology, and the Ultraviolet Imaging Telescope (UVIT) onboard the Indian AstroSat mission has given us unprecedented detail on extended objects such as planetary nebulae \cite{planetary} and supernova remnants (Crab, Vela, etc.). Understanding astrophysics – the physical conditions, temperatures, densities, and radiation fields – requires UV spectra of the atomic emission lines, which will complement the imaging observations. We have built the instrument,  we called the Spectroscopic Investigation of Nebular Gas (SING), to track emission lines over the entire spatial extent of a nebula. This is truly unique: previous observations have either mapped the spatial distribution of the emission or taken spectra at individual points, but never both simultaneously. 

SING is a long-slit UV spectrograph designed to operate in the NUV from 1400 \AA\, to 2700 \AA, with the primary goal to perform the spectroscopic survey of diffuse regions in the sky, such as supernova remnants, ISM, and planetary nebulae. When SING is mounted on a stable space platform in orbit, it will create a spectral survey of the sky in the NUV region.  Imaging surveys (such as e.g. GALEX and UVIT) have mapped the sky with high spatial resolution but without the spectral diagnostics necessary to probe local physical conditions. Our long-slit spectrograph SING will take simultaneous spectra of multiple locations to track different phases of the gas in extended regions from nebulae to galaxy clusters. The spectrograph was assembled, calibrated, and tested at the M.~G.~K.~Menon Laboratory (MGKML) for Space Sciences, located in  CREST campus of the Indian Institute of Astrophysics (IIA). This facility was used to integrate, characterise and calibrate the UVIT instrument onboard AstroSat \cite{amit}, and VELC onboard the ADITYA L1 missions \cite{venkata2022spectropolarimetry}. 

SING is accepted to be deployed on the Chinese modular space station (CSS) by the United Nations Office for Outer Space Affairs (UNOOSA) as part of the international cooperation for the utilization of the CSS for outer-space experiments. There has not been a UV astronomical experiment on a space station so far, except for the Glazar UV imaging telescope onboard the Mir Space Station, however, its sensitivity was lower than expected and it could only image the brightest stars in the UV \citep{glazar}. The space station provides a perfect stable platform for the spectroscopic survey over a long time period, and as the station orbits the Earth, SING will perform the spectroscopic survey of the areas of the sky determined by the orbital constraints. The data obtained from the mission will be made freely available to the public for further studies.

\section{Science Objectives}

\begin{figure}[ht]
\centering
  \begin{tabular}{@{}cc@{}}
    \begin{subfigure}[b]{0.49\textwidth}\includegraphics[width=\textwidth]
    {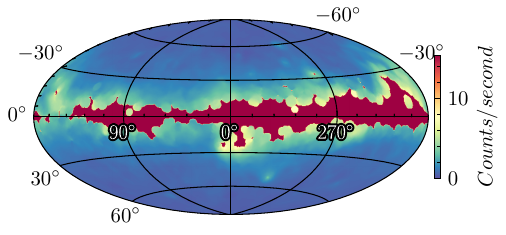} \caption{}\end{subfigure} &
    \begin{subfigure}[b]{0.49\textwidth}\includegraphics[width=\textwidth]{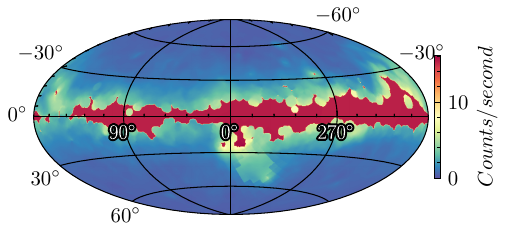}\caption{}\end{subfigure}  \\
    \begin{subfigure}[b]{0.49\textwidth}\includegraphics[width=\textwidth]{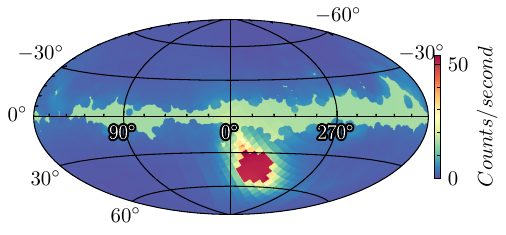}\caption{}\end{subfigure} &
    \begin{subfigure}[b]{0.49\textwidth}\includegraphics[width=\textwidth]{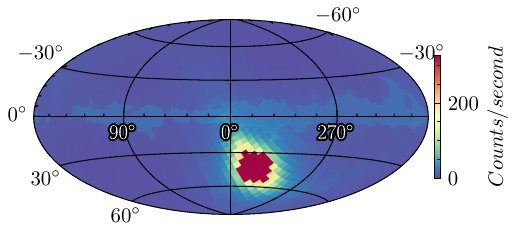}\caption{}\end{subfigure}   
\end{tabular}
\caption{UV background at 1400 \AA (a), 1800 \AA (b), 2200 \AA (c), and 2700 \AA (d) for a bandwidth of 10 \AA.}
  \label{back_sim}
\end{figure}

The primary objective for SING is to map the sky using the NUV ($1400 – 2700$ Å) long-slit spectrograph. Our primary targets are extended nebulae, including supernova remnants and planetary nebulae in our own Galaxy, star formation in nearby galaxies along with emission from their extended halos, and, on an even larger scale, the cosmic web \cite{Martin_2014}. Fig.~\ref{back_sim} shows the simulated UV background at 1400 \AA, 1800 \AA, 2200 \AA\ and 2700 \AA\ for a bandwidth of 10 \AA\ using {\it ASTUS} UV sky simulator \cite{astus} for a bin size of 0.5°/pixel. There are a number of important emission lines from many different species in this spectral region (Table~\ref{line}). For example, CIV (1550 Å) emission traces hot gas in supernova remnants and, along with other emission lines, is a sensitive diagnostic of the energetics and the dynamics in these regions. At the other end of the temperature scale, molecular hydrogen (H$_{2}$) emits in the Lyman and Werner bands in this spectral range, which will allow us to map the cold gas where much of the mass of the ISM in our Galaxy lies. Other emission lines give insight into the composition of the interstellar gas and their formation. Extragalactic objects have been well-mapped by imaging surveys like GALEX, but these only trace the morphology and not the astrophysics, i.e. the physical conditions, temperatures, densities, and radiation fields. The long-slit spectrograph will track hot gas and other phases from the interior of galaxy clusters to the cosmic web. 

The International Ultraviolet Explorer (IUE) Mission \cite{boggess1978iue} is seen as the precursor to all UV spectroscopic space missions. IUE operated in 1120 \AA\ to 3250 \AA\ wavelength range, focusing on observations of a wide range of objects from planets, stars, and nebulae to galaxies. IUE was followed by the Far Ultraviolet Spectroscopic Explorer (FUSE) \cite{FUSE,iue_nasa}, which operated in $900 - 1200$ \AA\ range, 
and Cosmic Origin Spectrograph (COS) \cite{cos} onboard the Hubble Space Telescope (HST) operated in $900 - 3200$ \AA\ range. The main science goals were to understand the origin of large-scale structures and the intergalactic medium, to study the formation and evolution of galaxies and to understand the origins of stellar and planetary systems. The recent Colorado Ultraviolet Transit Experiment (CUTE) mission is an NUV spectroscopic CubeSat mission focusing on the study of composition and mass loss of transiting hot Jupiter-like exoplanets \cite{brian}. The upcoming World Space Observatory (WSO) is a 1.7-m telescope which feeds a spectrograph with two Echelle channels and a long-slit spectrograph (LSS) with better than $0.5^{\as}$ resolution \cite{wso_shu,wso} but with smaller FOV than SING. The LSS will complement our instrument in such a way that we will be able to follow up the wide-field observations with more detailed maps from the WSO LSS. Any observations in a previously unexplored area have the potential to yield completely new science.

\begin{table}[h!]

\caption{List of atomic/molecular emission lines in SING range.}
\begin{tabular}{ll}
\hline
NUV lines & Peak \\
\hline                                      
OIII         & 2321, 2331                      \\
HeII         & 2307, 2386, 2422, 2512, 2734    \\
CIII         & 2297                            \\
NeIV         & 2423                            \\
CII          & 2326, 2836                      \\
OII          & 2470                            \\
Graphite     & 2175                            \\
NII          & 2140                            \\
CIII         & 1909                            \\
SiIII        & 1892                            \\
NIII         & 1750                            \\
AlII         & 1671                            \\
CIV          & 1548, 1551                      \\
SiII         & 1533                            \\
H2           & $1430-1620 $                     \\
\hline
\end{tabular}
\label{line}
\end{table}

\section{SING instrument design and overview}

The SING instrument design was done while keeping the wide FOV ($1\deg$) long slit requirement in mind for this spectroscopic survey. The preliminary conceptual design and science goals for SING were presented earlier in \cite{chandra2020spectroscopic,SING_SOS}. The instrument was designed to make use of the CSS platform, with a size constraint of $500 \times 500 \times 600$ mm. The instrument specifications are provided in Table~\ref{specs}.

SING mirrors are coated with Al and a protective layer of MgF$_2$ for good reflectivity in the 1400 -- 2700 \AA\ band. The detector is a Photek\footnote{\tt{http://www.photek.com/}} MCP with a caesium telluride (CsTe) solar-blind photocathode having QE $>10\%$ in the operating band, and a photon event processing unit consisting of CMOS camera and Raspberry Pi single board computer. 
To minimize the cost, we have manufactured the mirrors in-house at IIA. The spectrograph unit uses the slit, manufactured at the nanoscience fabrication facility of the Indian Institute of Science (CeNSE\footnote{\tt{http://www.cense.iisc.ac.in/}}, IISc, Bangalore) \cite{AS102-86}, and the holographic grating, manufactured by Horiba\footnote{\tt{https://www.horiba.com/int/scientific/technologies/diffraction-gratings/}}. The hardware design was optimized to meet the strict mass and volume limits specified by the launch provider. SING is designed to operate within a temperature range of $25 \pm 5 \deg C$, which will be ensured by using heaters and multi-layer-insulation (MLI).
\begin{table}[h!]
\centering
\caption{SING instrument basic parameters}
\begin{tabular}{lc}
\hline
FOV (L $\times$ W) & 1\si{\degree} $\times$ $3.5^{\prime\prime}$   \\
Primary aperture & 298 mm\\
Focal ratio & F/6.9  \\
Passband & $1400 - 2700$ \si{\angstrom }  \\
Plate scale & 1.69 arc minute/mm \\
Spectral Resolution  &  3.7 \si{\angstrom }  \\
Spatial Resolution &  $1.33^{\prime}$ \\
Peak Effective area  &  2.4 cm$^{2} $ \\
Mass & $<25$ kg \\
Instrument size  & 510 $\times$ 400 $\times$ 400 mm \\
Detector & Solar-blind MCP\\
Pore size & 10 $\mu$m \\
Time resolution & 33 ms \\
\hline
\end{tabular}
\label{specs}
\end{table}

\subsection{Optical design}

SING optical unit is divided into two parts: the telescope and the spectrograph. The telescope uses a basic parabolic primary mirror and a spherical secondary mirror, with a tilt mirror to focus and steer the beam to the spectrograph. The spectrograph has an entrance slit, a concave holographic grating, and an MCP detector. The slit was designed in a dumbbell shape for a field of $1.1 \deg$. The advantage of a dumbbell slit is that where the resolution degrades in the extreme fields, the width of the slit is increased to let more light through without affecting the resolution. The advantage of concave grating is that it acts as both a dispersing and focusing element, reducing the number of reflections without the need for collimating and focusing optics, as each reflection costs nearly $20\%$ of the photons. The main goal of this overall instrument design was to reduce the cost of development, assembly and testing. In addition, we could manufacture some of the components, such as mirrors, locally at the IIA Photonics Lab\footnote{\tt{https://www.iiap.res.in/?q=photonics.htm}}.
\begin{figure}[ht]
    \centering
\includegraphics[width=0.8\linewidth]{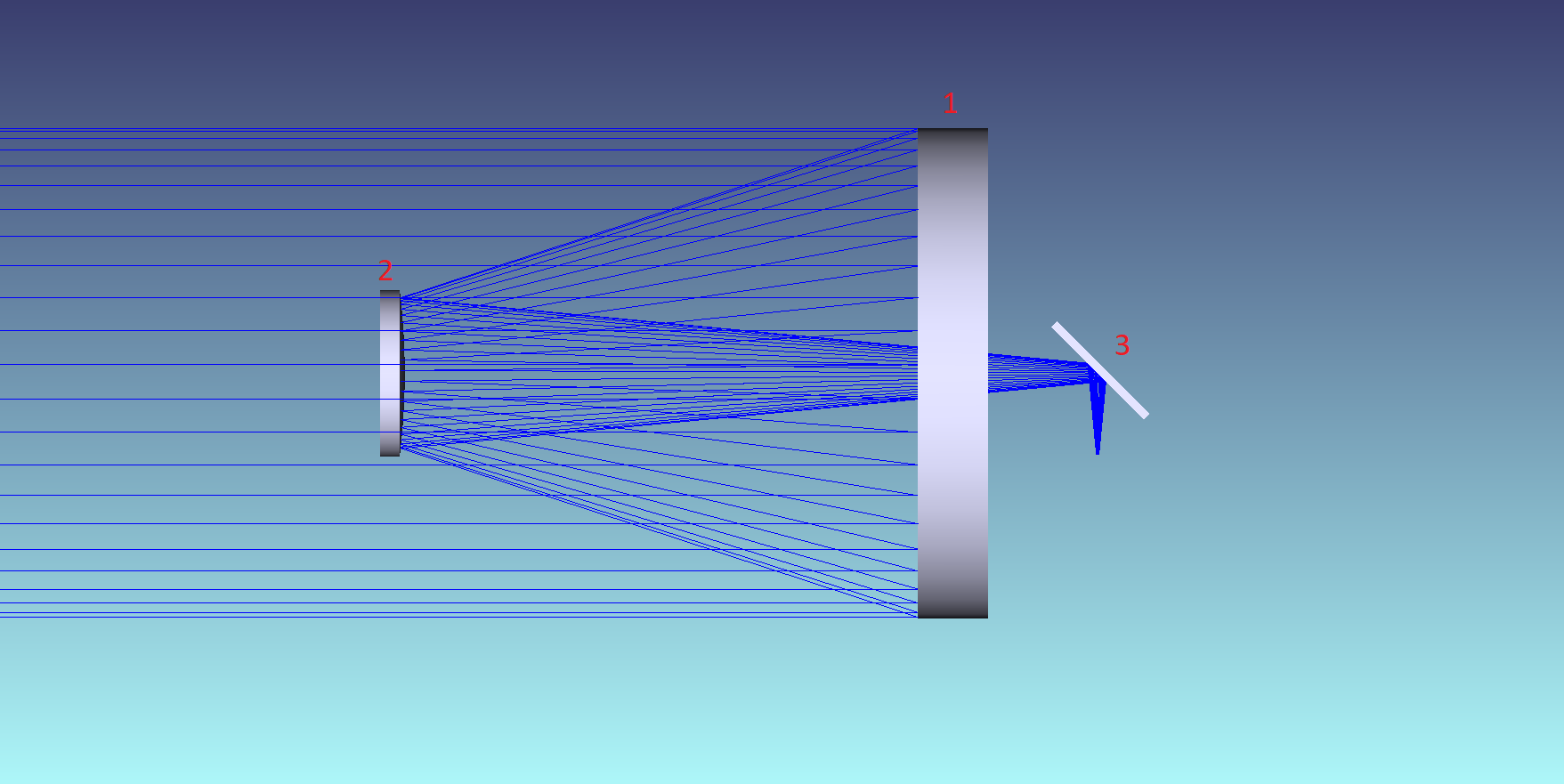}
    \caption{Zemax optical layout of the telescope design: 1. primary mirror; 2. secondary mirror; 3. tilt mirror.}
    \label{fig:tele_layout}
\end{figure}

\begin{table}[h!]
\centering
\begin{tabular}{lllcll}
\hline
  Surface & Radius of curvature & Thickness& Conic (k) & Material & Size (mm)  \\
 \hline
 Primary Mirror & 937.5 mm& 50 mm & -1&  BK7    & 298 dia\\
 Secondary mirror & 416.88 mm& 15 mm&     0              & Float glass & 104 dia\\
 Tilt Mirror &  inf & 10 mm & 0& Zerodur                                          & 60 x 19\\
 \hline
\end{tabular}
\caption{Telescope optical parameters}
\end{table}

The design was developed using Zemax\footnote{http://www.zemax.com/}, and parameters like the radius of curvature, the distance between optical surfaces, and conic constants were optimised to get the best spot size for the telescope and the spectrograph. Fig.~\ref{fig:tele_layout} shows the optical layout of the SING telescope. The parallel beam of light, coming from the source, is focused by the parabolic f/1.57 primary mirror with a radius of curvature of 937.5 mm onto the concave secondary mirror with a radius of curvature of 416.88 mm. The secondary mirror focuses the rays at the back focal point. The final f-ratio of the telescope is f/5, with an effective focal length of 1500 mm and a plate scale of 2.29 arc minute per mm. A tilt mirror is placed at $45 \deg$ to tilt the beam toward the spectrograph slit. 
\begin{figure}[h]
    \centering
\includegraphics[width=0.6\linewidth]{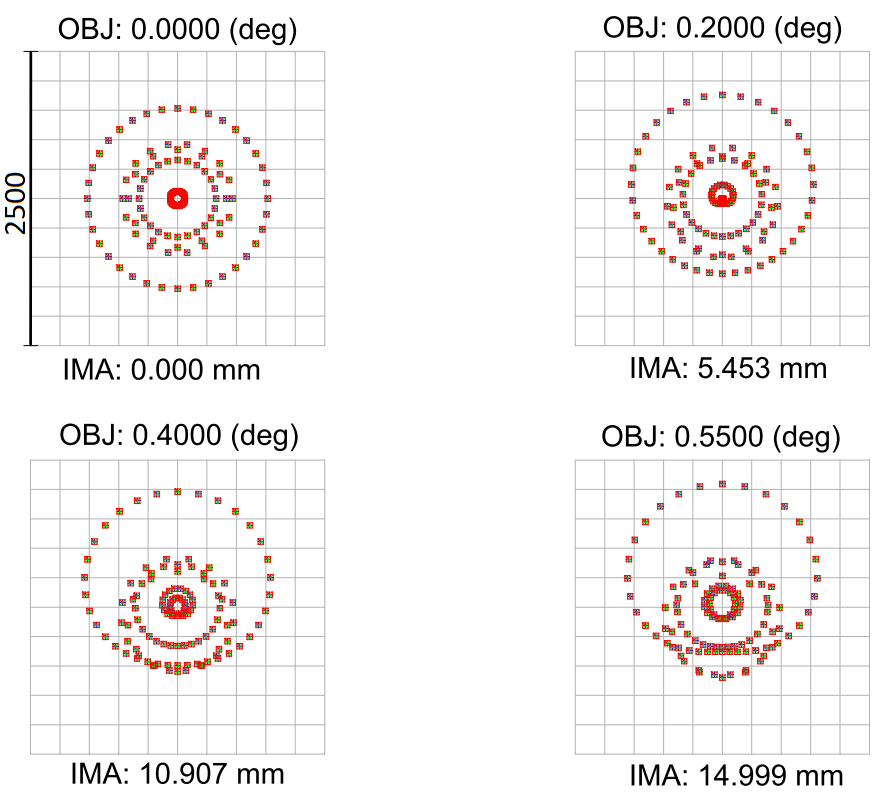}
\vspace{2mm}
\caption{Spot diagram of telescope at field angle $0\deg$, $0.2\deg$, $0.4\deg$ and $0.55\deg$ for wavelengths 2400 \AA\ (blue), 2000 \AA\ (green), and 2700 \AA\ (red). }
\label{fig:spot_diagram}
\end{figure} 

The primary mirror is made from BK7 optical glass, the secondary from float glass, and the tilt mirror from Zerodur\textsuperscript{\textregistered}, all of them grounded, polished and tested at the IIA photonics lab. The reflective surfaces were coated\footnote{HHV: Hind High Vacuum Company Private Ltd: https://hhv.in/} with reflective Al coating and a protective MgF2 layer \cite{coating,coating1}, providing reflectivity greater than 85\% in the operating NUV band. The MgF2 layer prevents Al from oxidising, as even a few nm-thick oxide layers can degrade the reflectivity of the Al coating in the UV. The telescope has an RMS spot radius of 498 $\mu m$ at the centre field, and 516 $\mu m$ at the $0.55\deg$ edge field (Fig.~\ref{fig:spot_diagram}). To prevent the straylight from entering the optical system, a baffle was designed following the method described in \cite{baffle}, to allow a field of $ \pm 0.56\deg$. The baffle was black anodised to reduce scattering on the baffle surface.

\begin{minipage}[t]{0.6\textwidth}
\begin{figure}[H]
    \centering
\includegraphics[width=1\linewidth]{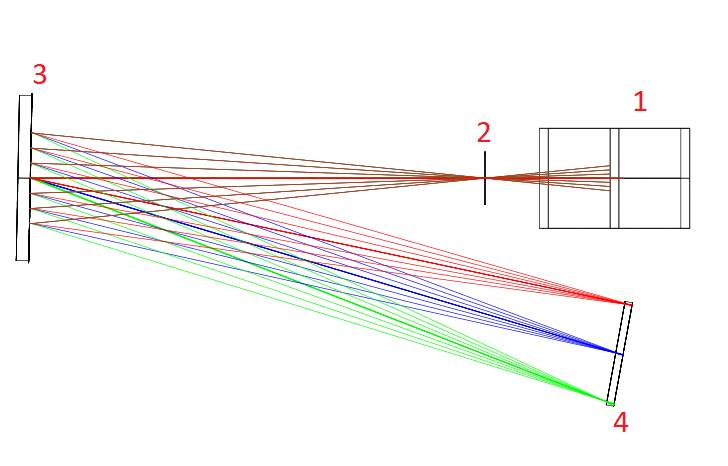}
\vspace{7mm}
    \caption{Zemax optical layout of the spectrograph design: 1. tilt mirror; 2. dumbbell slit; 3. holographic concave grating; 4. detector.}
    \label{fig:spec_layout}
\end{figure}
\end{minipage}
\hfill
\begin{minipage}[t]{0.35\textwidth}
\begin{figure}[H]
    \centering
     \includegraphics[width=1\linewidth]{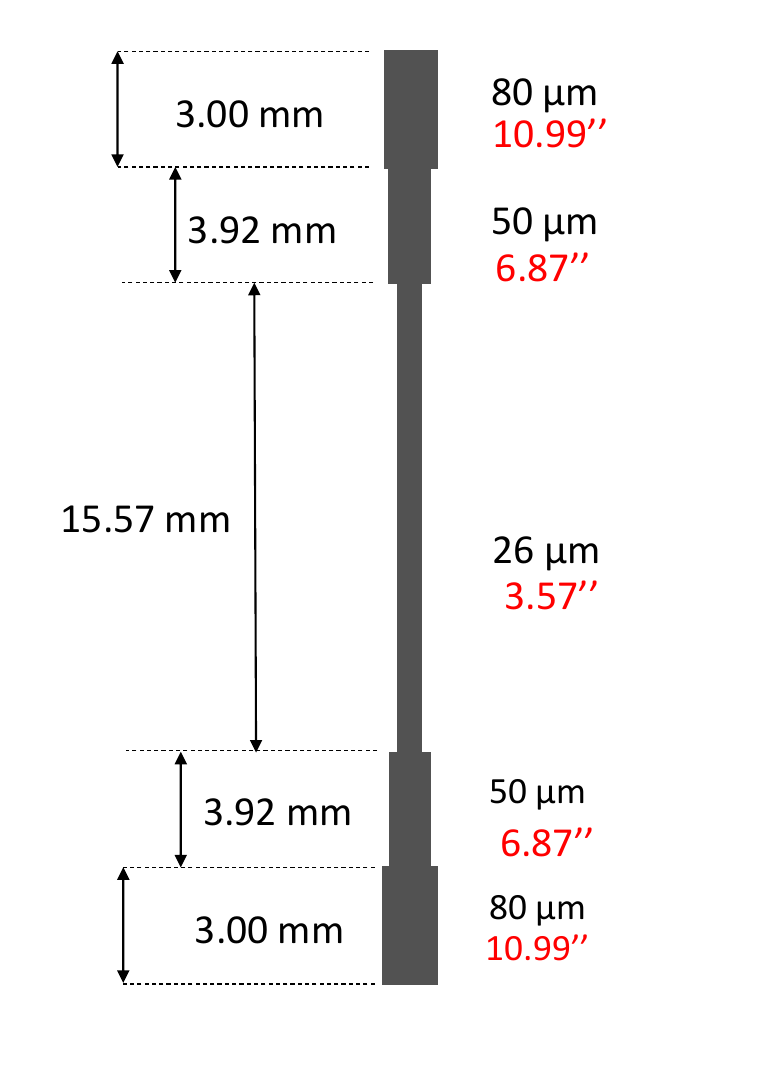}
    \caption{Dimensions of the slit.}
    \label{fig:slit}
\end{figure}
\end{minipage}

Fig.~\ref{fig:spec_layout} shows the optical layout of the spectrograph. The concave holographic grating of 75-mm diameter has a focal length of 201.06 mm and a dispersion of 3.25 nm/mm. The concave grating images the slit at different wavelengths onto the 40-mm diameter MCP detector, covering wavelengths of 1400 \AA\, to 2700 \AA\, with a resolution $3.7$ \AA\, and a field of $1.1\deg$. Here the concave grating acts as the dispersing and focusing element. Fig.~\ref{fig:slit} shows the dimensions and the shape of the slit with the field marked for reference. The specifications of the grating and slit used are provided in Table~\ref{spectro_ref}. The final number f-ratio at the detector is f/6.9, and the plate scale is 1.69 arc minute/mm. Since the spot size of the telescope is much bigger than the slit width of 25 $\mu$m, the slit efficiency is only 10\%, but because we are mainly focusing on diffuse sources, this is sufficient for our purposes.

\begin{table}[h!]
\centering
\begin{tabular}{ll}
\hline
\multicolumn{2}{c}{\textbf{Grating}} \\
\hline
Type  & Holographic \\
    &flat-field corrected\\
Material & Fused silica (dia. 75 mm)\\
Grating surface & Concave\\
Radius of curvature & 207.06 mm\\
Groove density & 1200 lines/mm \\
Dispersion  &   3.25 nm/mm \\
\hline
\multicolumn{2}{c}{\textbf{Slit}} \\
\hline
Type & Dumbbell \\
Substrate & Silicon Wafer \\
Length & 30 mm\\
Center width & 26 $\mu$m\\
Substrate Thickness & 100 $\mu$m\\
\hline
\end{tabular}
\caption{Spectrograph parameters}
\label{spectro_ref}
\end{table}

\subsection{Mechanical design}

\begin{figure}[h!]
\includegraphics[width=0.5\textwidth]{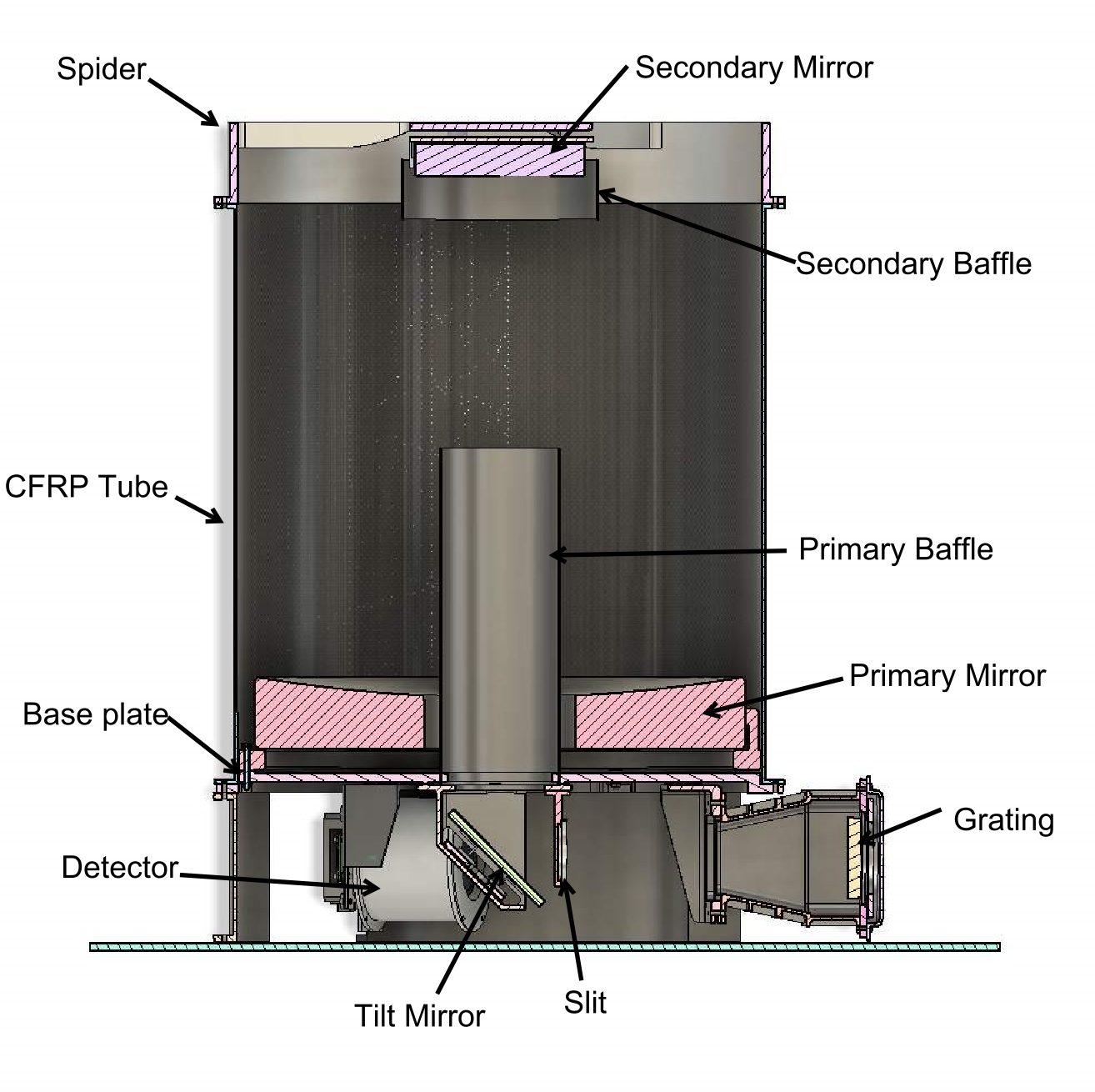}
\hskip 0.3in
\includegraphics[width=0.4\textwidth]{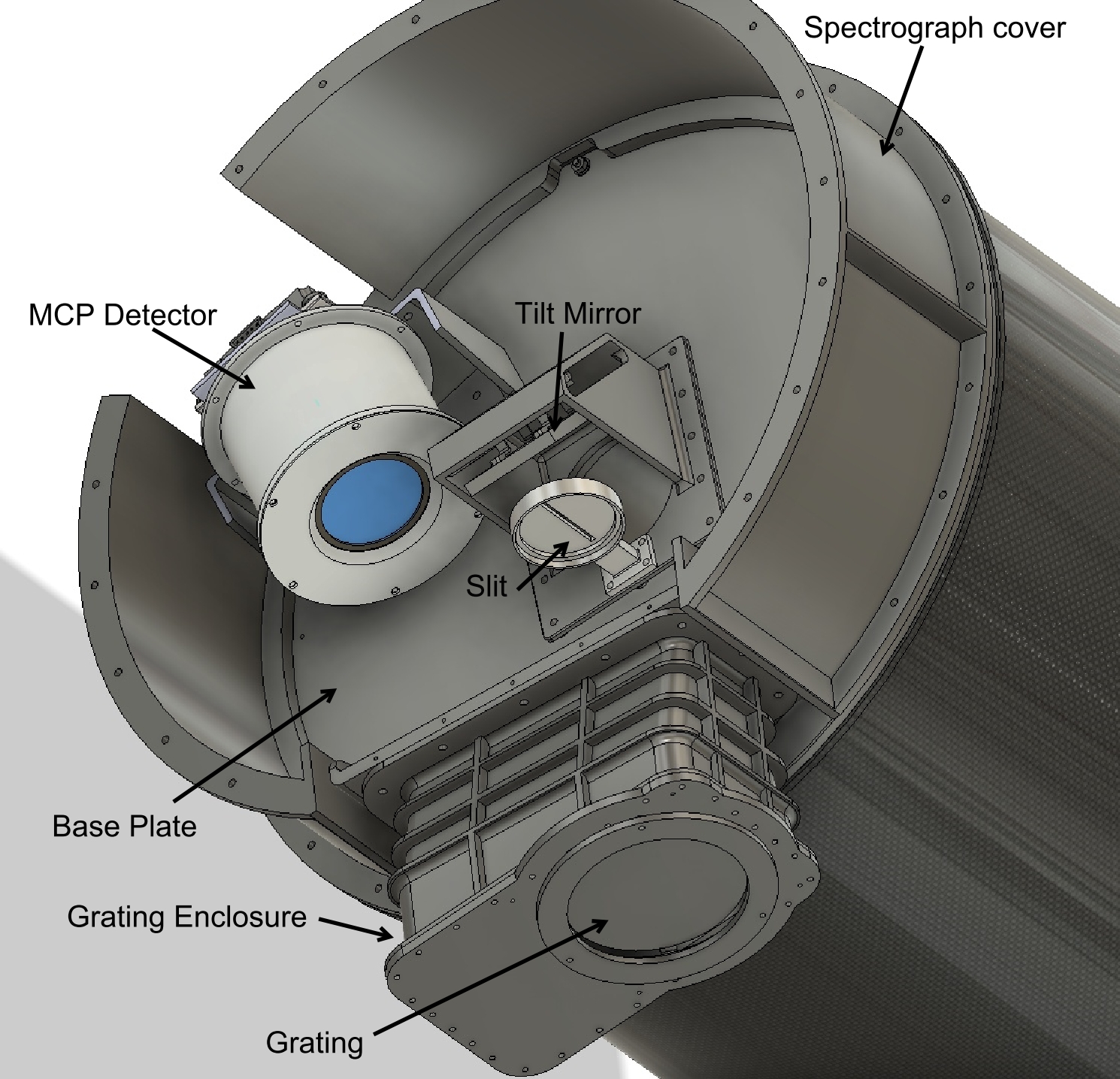}
\caption{{\it Left}: CAD rendering of SING cross-section. {\it Right}: CAD rendering of the SING spectrograph.}
\label{fig:sing_cross_section}
\end{figure}

The instrument's mechanical structure was designed in accordance with stringent requirements such as size, mass, launch loads, and thermal stability in mind. The launch requirements were based on Long March-2F manned spaceship\footnote{https://www.cnsa.gov.cn/english/n6465715/n6465718/c6476863/content.html} flight with a longitudinal root-mean-square acceleration of 6.5 G$_{\rm rms}$, and lateral of 3 G$_{\rm rms}$. The mechanical structure of SING was made from Al 6061-T6 and CFRP (Carbon Fiber Reinforced Polymer)\footnote{UNICITA Consulting Pvt Ltd.}. The primary structure of the telescope subsystem consists of the metering tube made from MJ46 composite CFRP, primary mirror mount, secondary spider assembly, base plate assembly with mounts, and protective cover made from Al~6061-T6 (Fig.~\ref{fig:sing_cross_section}). The secondary spider assembly houses the secondary mirror with the mirror mount and a baffle. The base plate holds the CFRP tube and the primary mirror, with the mirror mount on one side and the spectrograph components on the other side. 

The spectrograph is housed behind the base plate and consists of the slit mount, grating assembly, and detector assembly, all made from Al 6061-T6. Fig.~\ref{fig:sing_cross_section} shows the cross-section view of the SING structural components. Both telescope and spectrograph subsystems were designed following the thermal requirements of the launch provider.

\subsection{Detector system}

SING UV detector operates in the photon-counting mode as UV flux from most sources is low. In this mode of operation, each photon is tagged with X,Y position \cite{MICHEAL} and time. This detector consists of a 40-mm dia $Z$-stack MCP manufactured by Photek operating at a high gain of $10^7$. The MCP has a solar-blind photocathode made of CsTe sensitive to UV and a maximum quantum efficiency (QE) of about 20\% in the operating band. The MCP is powered by a high voltage power supply FP632\footnote{ Supplied along with the MCP.}, which provides the required $5500$ V, $2400$ V and $-200$ V voltages for the MCP operation. 

We have developed a readout and photon event processing board comprising Raspberry Pi (RPi) cam V2 as the camera and a Raspberry Pi Zero 2W\footnote{\tt{https://www.raspberrypi.org/}} as an embedded microprocessor platform \cite{pcd}.  Here, we take advantage of the inbuilt readout implemented in the RPi Graphics Processing Unit (GPU). The centroid calculation of the photon events is carried out by the Raspberry Pi Zero 2W processor. We have also implemented single-event recovery and system reset in case of a malfunction in the detector system. The camera is a Sony IMX 219, 8-MP camera operating in $1280\times720$-pixels video mode at 30 frames per second with a fixed exposure time of 33 ms. The phosphor screen behind the MCP is imaged by this camera
and these images are processed in real-time to detect events and perform centroiding using the $5\times5$ window algorithm \cite{S.Ambily}. In this mode of operation, to detect a photon event, a $5\times5$ window scans for a central pixel value above a threshold and a window sum value above an energy threshold; when this condition is satisfied, centroiding is performed. We tag the photon events with the frame number and the maximum and minimum corner pixel values in the $5\times5$ windows. The maximum and minimum corner pixel values are used to detect double events \cite{hutchings}. Figs.~\ref{fig:cross_section_detector} and \ref{fig:sing_detector_assembly} show the SING detector assembly. We found that the detector has a dark count of 68 counts per second\footnote{Detector was covered, and counts were measured for 3 min. We found the counts to be around 68 counts per second, which matched with the value of 70 counts per second provided by Photek.}. We do not expect the dark counts of the Photek detector to be different from the ground values, but cosmic ray interaction can increase the count rate, which needs to be analysed from the in-orbit data\cite{Tandon_2017}.The detector parameters are given in table~\ref{detector_param}.

\begin{minipage}[t]{0.5\textwidth}
\begin{figure}[H]
    \centering
    \includegraphics[width=\linewidth]{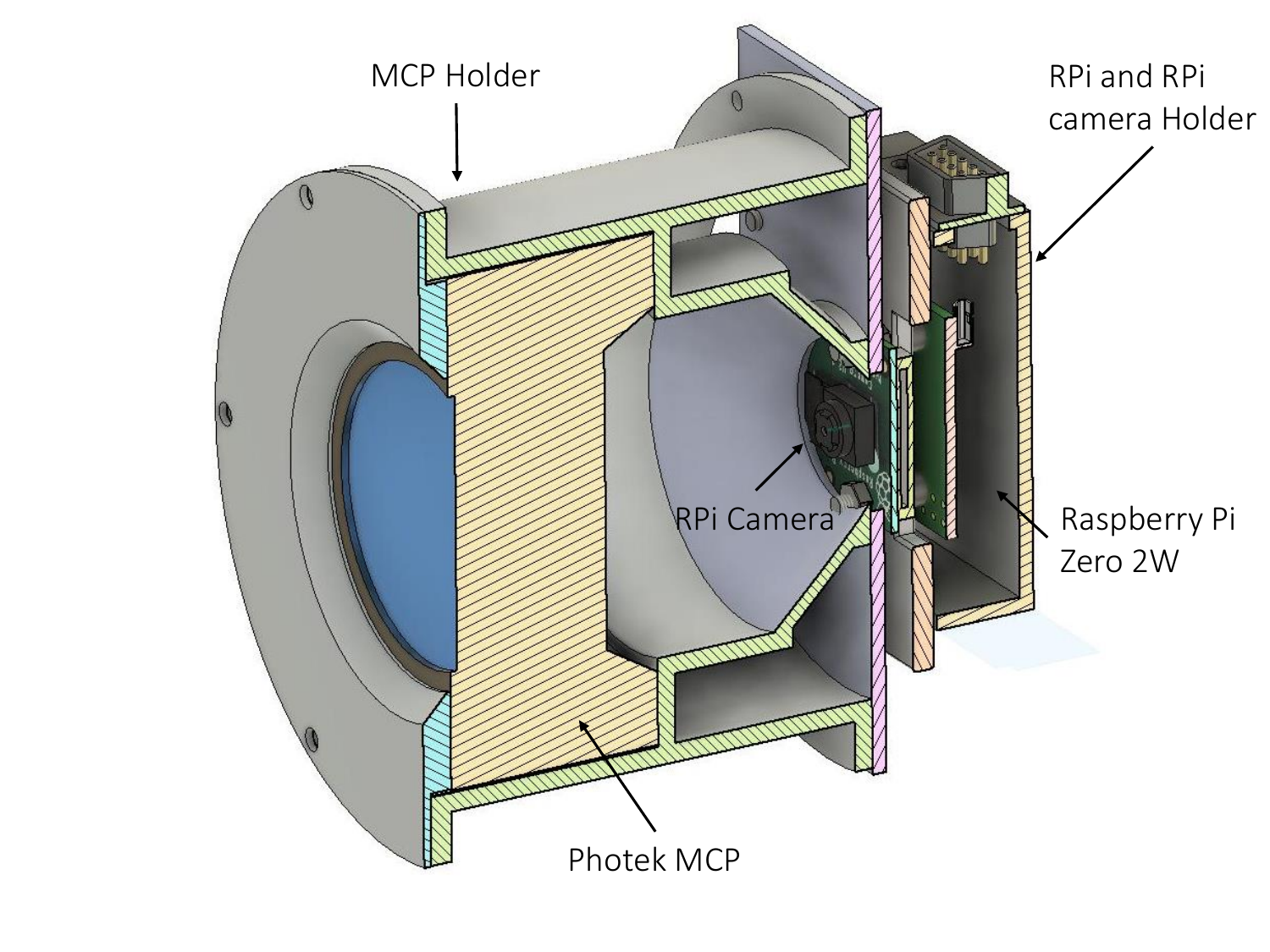}
    \caption{CAD layout of SING detector cross section.}
    \label{fig:cross_section_detector}
\end{figure}
\end{minipage} 
\hfill
\begin{minipage}[t]{0.45\textwidth}
\begin{figure}[H]
    \centering
    \includegraphics[width=\linewidth]{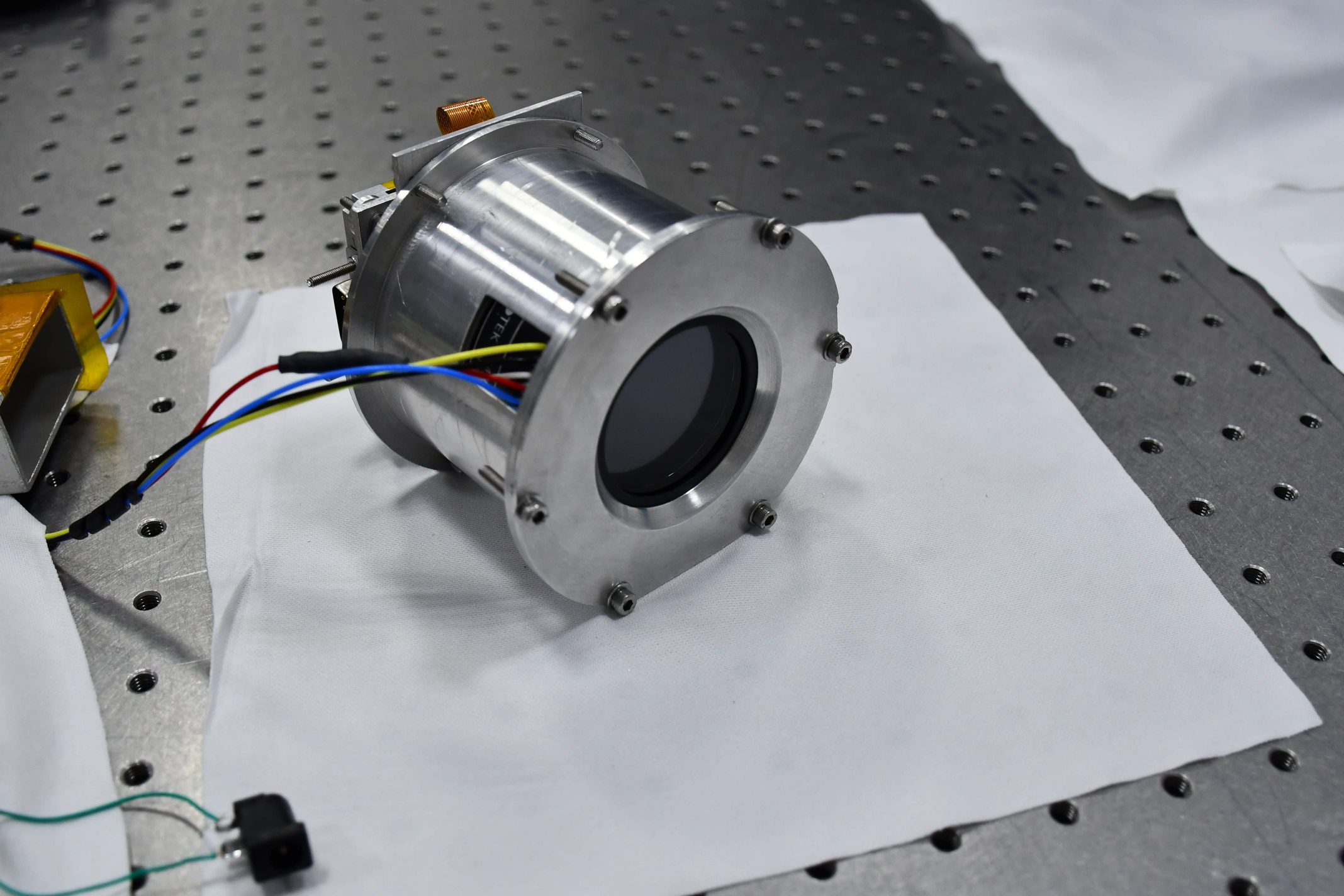}
    \caption{Assembled SING detector.}
    \label{fig:sing_detector_assembly}
\end{figure}
\end{minipage}

\begin{table}[h!]
\centering
\begin{tabular}{ll}
\hline
\multicolumn{2}{c}{\textbf{Detector Parameter}} \\
\hline
Detector & MCP \\
Photo cathode  & CsTe\\
Readout & CMOS camera + RPi Zero 2W\\
CMOS Sensor & Sony IMX219 \\
Operating Mode & Photon Counting\\ 
CMOS pixel format & 3280 $\times$ 2464\\
Capture Resolution & 1280 $\times$ 720\\
Time Resolution & 33 ms \\
Interface & UART\\
\hline
\end{tabular}
\caption{Detector parameters}
\label{detector_param}
\end{table}

\section{Instrument alignment and assembly}

\begin{figure}[H]
\centering
\sbox{\measurebox}{%
  \begin{minipage}[b]{.6\textwidth}
  \subfloat
    []
    {\label{fig:figA}\includegraphics[width=\textwidth]{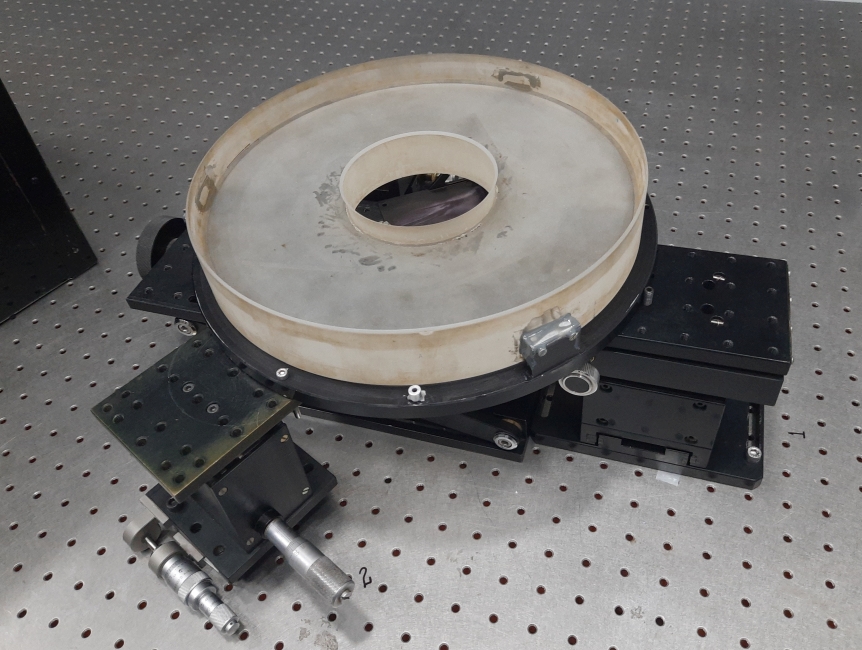}}
  \end{minipage}}
\usebox{\measurebox}\qquad
\begin{minipage}[b][\ht\measurebox][s]{.25\textwidth}
\centering
\subfloat
  []
  {\label{fig:figB}\includegraphics[width=\textwidth]{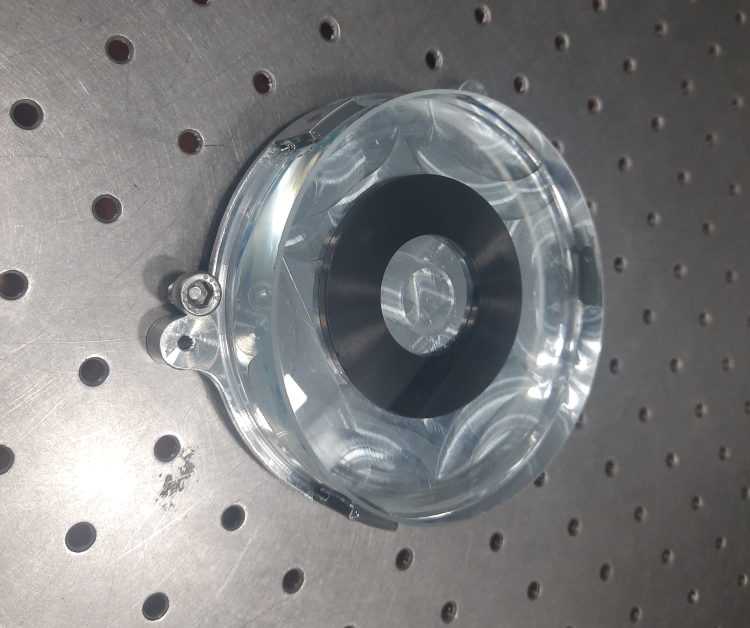}}

\vfill

\subfloat
  []
  {\label{fig:figC}\includegraphics[width=\textwidth]{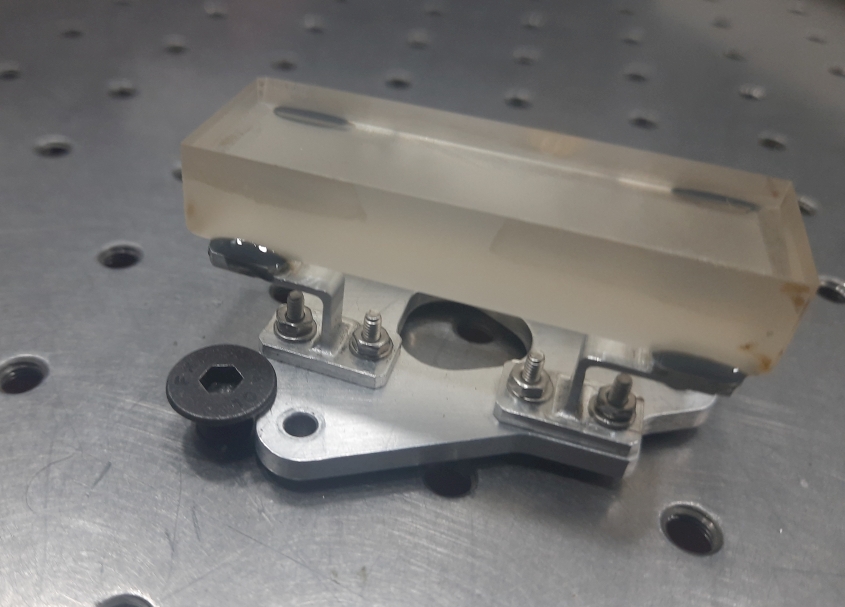}}
\end{minipage}
\caption{Mirrors bonded to their mounts. (a) primary, (b) secondary and (c) tilt mirror, respectively.}
\label{fig:glueing}
\end{figure}

The first procedure for the assembly was to bond the three mirrors (primary, secondary, and tilt) onto their respective mounts, before applying the UV reflective coating to reduce contamination. For bonding, the primary mirror mount was placed on optical jacks on an optical table, and the flatness of the mount was verified with a level gauge to an accuracy of $0.004\deg$. The primary mirror was then rested on another optical jack, with a height more significant than the primary mirror mount blades. 3M\textsuperscript{\footnotesize
{TM}} Scotch-Weld  2216 grey epoxy was applied to the blades, and the primary mirror was slowly lowered into the position, keeping a 2-mm gap between the mount base and the back of the mirror (Fig.~\ref{fig:glueing}). For the secondary and tilt mirrors, the mount flatness was first established. Then, using shims, the mirrors were slowly positioned in the mounts after applying the 2216 epoxy to the mount blades. Once the epoxy cured for a week, the mirrors with their mounts were sent for UV coating. 
    
\begin{table}[h!]
\centering
\caption{Alignment tolerance limit allowed for SING alignment.}
\begin{tabular}{llc}
\hline
 Tolerance term & objects  & tolerance\\
 \hline
  Decenter X \& Y ($\mu m$) & Primary mirror &  $\pm 50$ \\
                     & Secondary mirror & $\pm 50$ \\
                     & Tilt Mirror & $\pm 50$ \\
                     & Slit  &  $\pm 50$   \\
                     & Grating &$\pm 50$\\
 Tilt in X \& Y($^{\as}$)  & Primary mirror &  $60$ \\
                     & Secondary mirror & $60$ \\
                     & Tilt Mirror & $60$ \\
                     & Slit  &  $60$   \\
                     & Grating &$60$\\
                     \hline 
\end{tabular}
\end{table}

The grating was also bonded to the mount using a procedure similar to that for a secondary mirror (Fig.~\ref{fig:grat_glu}). The slit was bonded to the custom-made mount using 2216 epoxy; markings were made on the mount and the slit to ensure proper centering. Then the 2216 epoxy was deposited on the edges of the slit mount far away from the slit aperture (Fig.~\ref{fig:slit_glu}). The slit was slowly lowered on the mount while making sure the marking was close. Once lowered, using tweezers, the markings were perfectly aligned and set to cure for a week.

\begin{minipage}[t]{0.4\textwidth}
\begin{figure}[H]
    \centering
    \includegraphics[width=\linewidth]{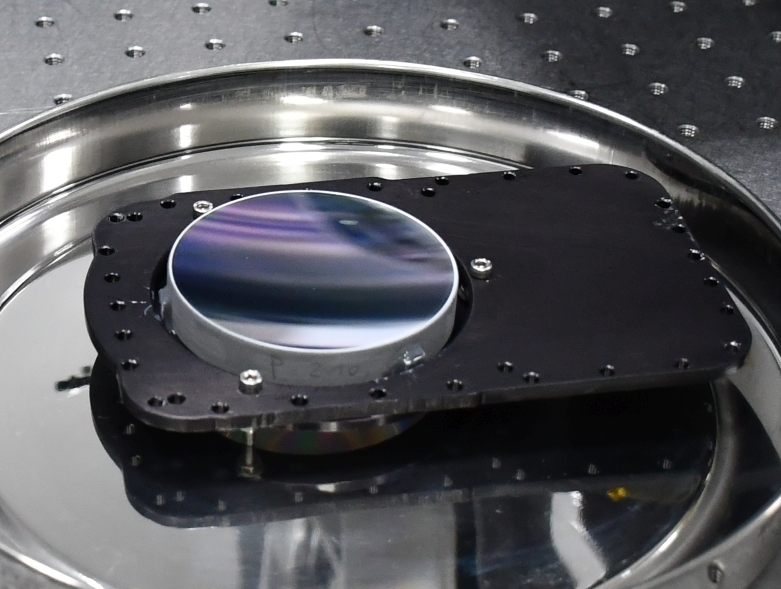}
    \caption{bonding of the grating on to grating mount.}
    \label{fig:grat_glu} 
\end{figure}
\end{minipage}
\hspace{10mm}
\begin{minipage}[t]{0.4\textwidth}
\begin{figure}[H]

    \centering
    \includegraphics[width=\linewidth]{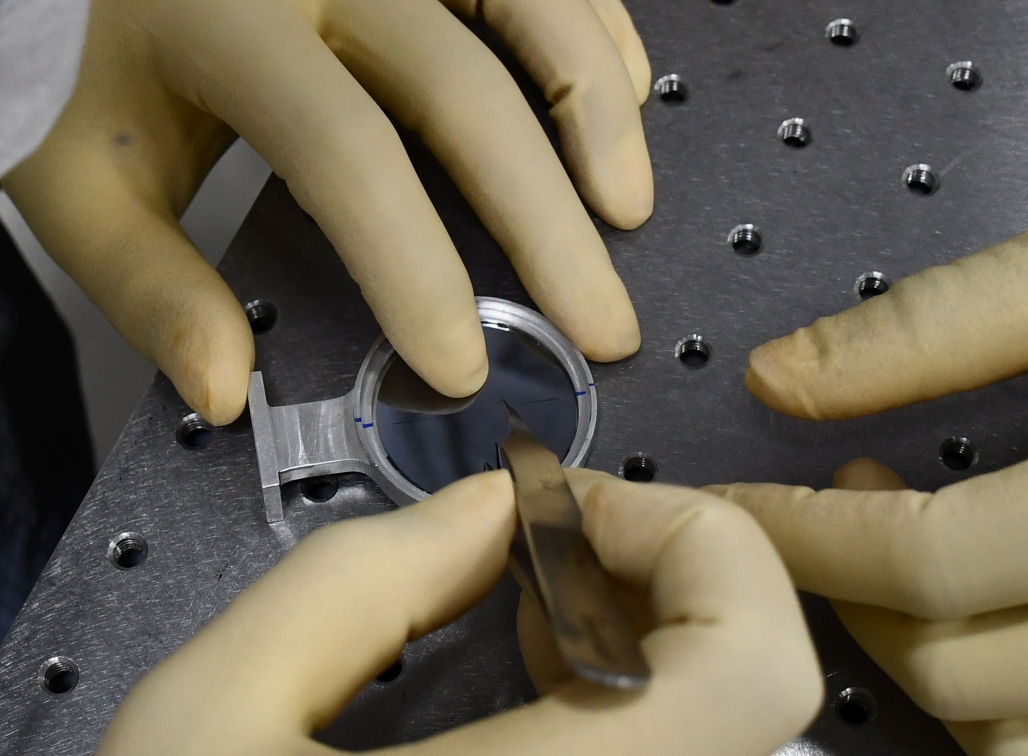}
    \caption{Gluing of the slit to the slit mount.}
    \label{fig:slit_glu}
\end{figure}
\end{minipage}
\subsection{Telescope alignment}

\begin{figure}[H]
    \centering
\includegraphics[width=0.9\linewidth]{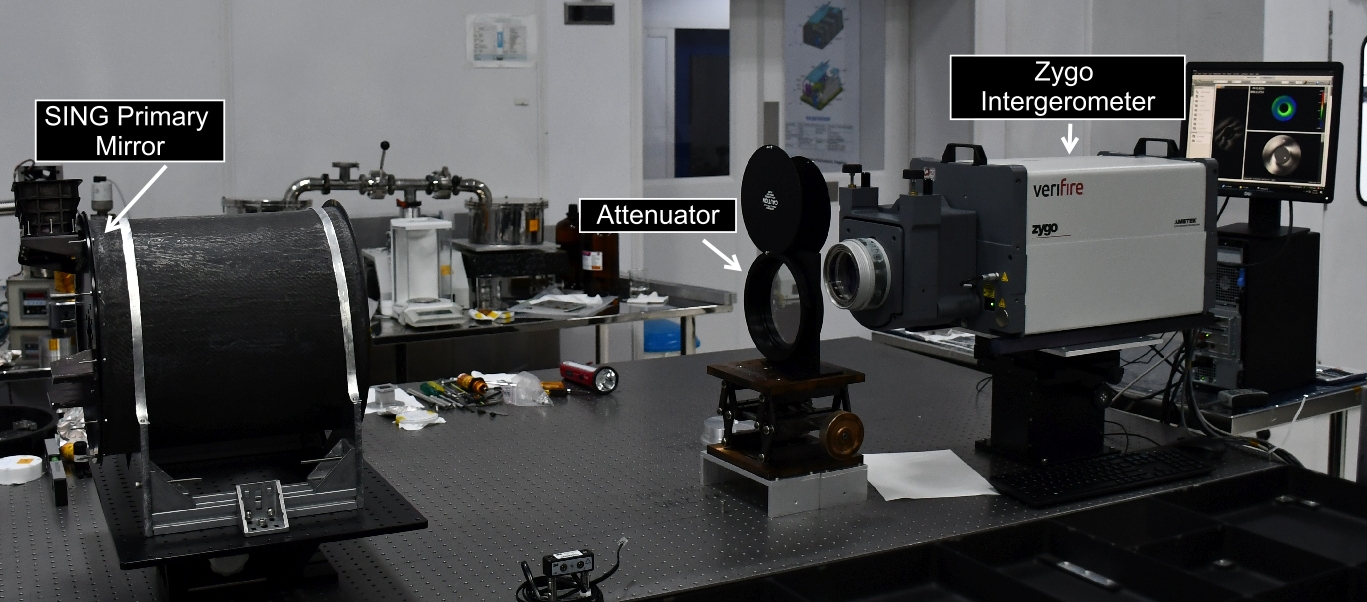}
    \caption{Test setup for primary alignment.}
    \label{fig:prim_setup}
\end{figure}

\begin{figure}[ht!]
\centering
\includegraphics[width=0.55\textwidth]{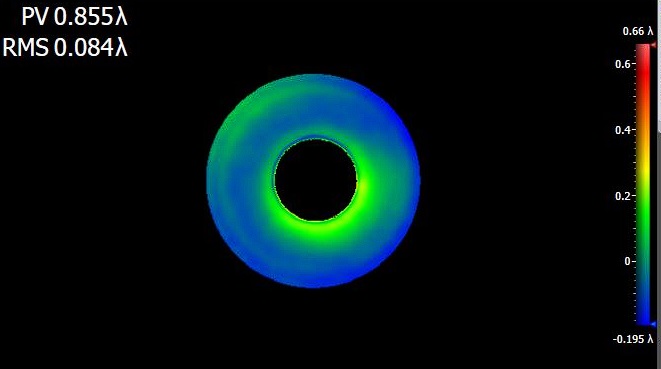}
\hskip 0.1in
\includegraphics[width=0.35\textwidth]{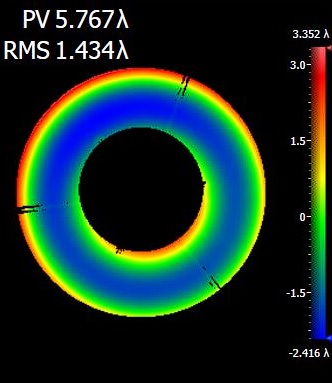}
\caption{{\it Left}: Primary mirror surface profile. {\it Right}: Combined surface profile.}
\label{fig:combined_zygo}
\end{figure}

The primary mirror with the mount was bolted onto the base plate, and the base plate was then mounted onto the CFRP telescope tube. This setup was rested on an optical jack collaring at the CFRP tube. A Zygo Verifire interferometer with an XYZ optical stage was mounted on an optical table (Fig.~\ref{fig:prim_setup} shows the test setup). Then, the optical axis of the Zygo was established. The optical axis of the primary mirror was aligned to the Zygo axis by operating the Zygo in alignment mode. Once the axis was aligned, we recorded the interferogram of the primary mirror to check for residual stresses in the mirror (Fig.~\ref{fig:combined_zygo}~{\it Left} shows the surface profile). Following this, the primary baffle was mounted on the base plate.

\begin{figure}[h!]
    \centering
\includegraphics[width=0.85\linewidth]{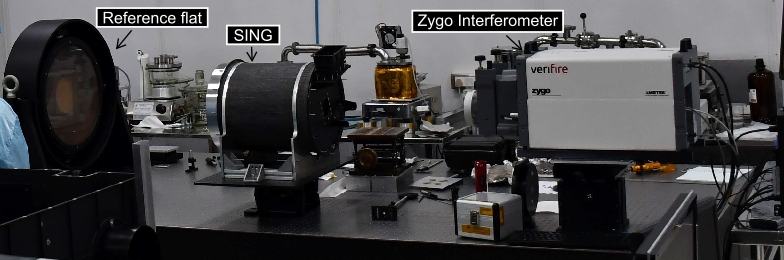}
    \caption{Test setup for SING telescope alignment.}
    \label{fig:tele_setup}
\end{figure}

The secondary mirror was then mounted on the spider along with its baffle. The base plate with the primary mirror and CFRP tube were positioned facing away from Zygo, as shown in Fig.~\ref{fig:tele_setup}, and a reference flat was placed in front of the primary mirror tube. Once the optical axis was established, the spider with the secondary mirror was mounted. The light from the f/3.3 sphere travels from the back focal plane of the telescope to the secondary mirror, is reflected onto the primary mirror, and travels towards the reference flat. The beam then retraces the path backwards. Initial adjustments were made to the spider in alignment mode, and fine adjustments were made using shims to reduce the effects of Zernike coefficients, mainly spherical term, decentre term, and defocus term. Once the alignment of these mirrors was completed, a combined interferogram was taken (Fig.~\ref{fig:combined_zygo}, {\it Right}). After this, the tilt mirror was added, and it was made sure that the focus lies on the slit plane on the slit mount, after which the slit was mounted.

\subsection{Spectrograph alignment}

\begin{figure}[H]
    \centering
\includegraphics[clip,trim={0 0.5cm 0 0.5cm},width=0.5\linewidth]{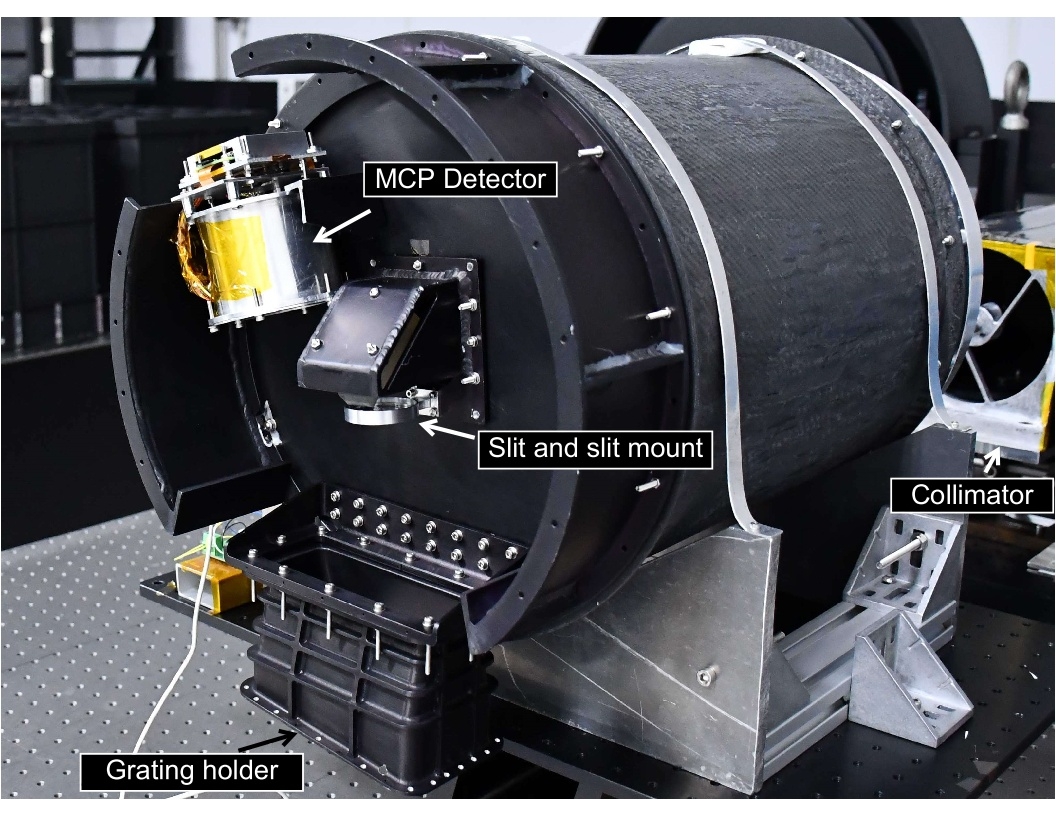}
    \caption{Test setup for SING spectrograph alignment.}
    \label{fig:spec_setup}
\end{figure}

To assemble and align the spectrograph, an f/4 600-mm Newtonian telescope was used as a collimator with an Acton VM-504 monochromator with a deuterium source as input. We illuminated SING with the collimator, and the monochromator was set to $0$-th order. It was made sure that the light was focused at the centre of the slit. The detector was then mounted on the detector holder on the base plate. Now, the bonded grating was attached to the grating holder, and the grating holder was mounted to the grating cover so that the continuum was visible at the centre of the detector. Then, the monochromator was set at 2700 \AA\ and, using shims to control the tilt and fine adjustment, the beam was directed to fall on the edge of the detector. The detector position was adjusted by fine movement using shims to get best possible focus for 2200 \AA. The assembled spectrograph setup is shown in Fig.~\ref{fig:spec_setup}. 

\section{Calibration}

\subsection{Effective area}

\begin{figure}[H]
    \centering
\includegraphics[width=0.7\linewidth]{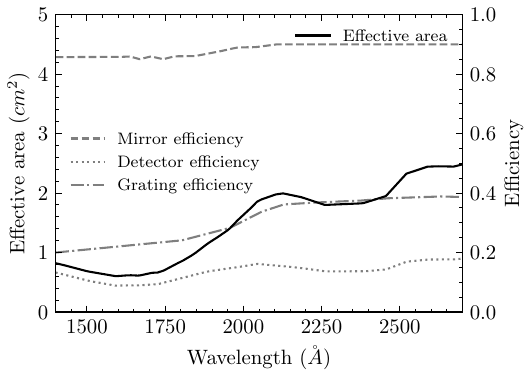}
    \caption{SING effective area as a function of wavelength.}
    \label{fig:effective_area}
\end{figure}

Total effective area of SING, $A(\lambda)$, is a function of wavelength($\lambda$) and is given by 
\begin{equation}
A(\lambda)=A_{T} M(\lambda)^3  S_{\eta}  G(\lambda)  \eta(\lambda) \,,
\label{eq:effective}
\end{equation}
where $A_{T}$, $P(\lambda)$, $S_{\eta}$, $G(\lambda)$, and $\eta(\lambda)$ are the effective geometrical collecting area of the telescope (612 cm$^{2}$), the reflectivity of mirrors as provided by the HHV, slit throughput measured in the lab (8\%), efficiency of holographic grating as provided by Horiba, and quantum efficiency of the MCP detector provided by Photek, respectively. Fig.~\ref{fig:effective_area} shows the calculated effective area of SING as a function of wavelength. The optical system has a maximum effective area of 3 cm$^2$ at 2700 \AA. The effective area at wavelengths 1400 \AA\ and 2000 \AA\ is 1 cm$^2$ and 2 cm$^2$, respectively. For a solid angle of $5.837 \times\ 10^{-7}$ steradians of the slit and an effective area of 1.5 $cm^2$, we found the instrument sensitivity to be around 0.069 $counts\ sec^{-1} Rayleigh^{-1}$\cite{counts_per}.  

\subsection{Spectral resolution}

\begin{minipage}[t]{0.5\textwidth}
\begin{figure}[H]
    \centering
    \includegraphics[width=\linewidth]{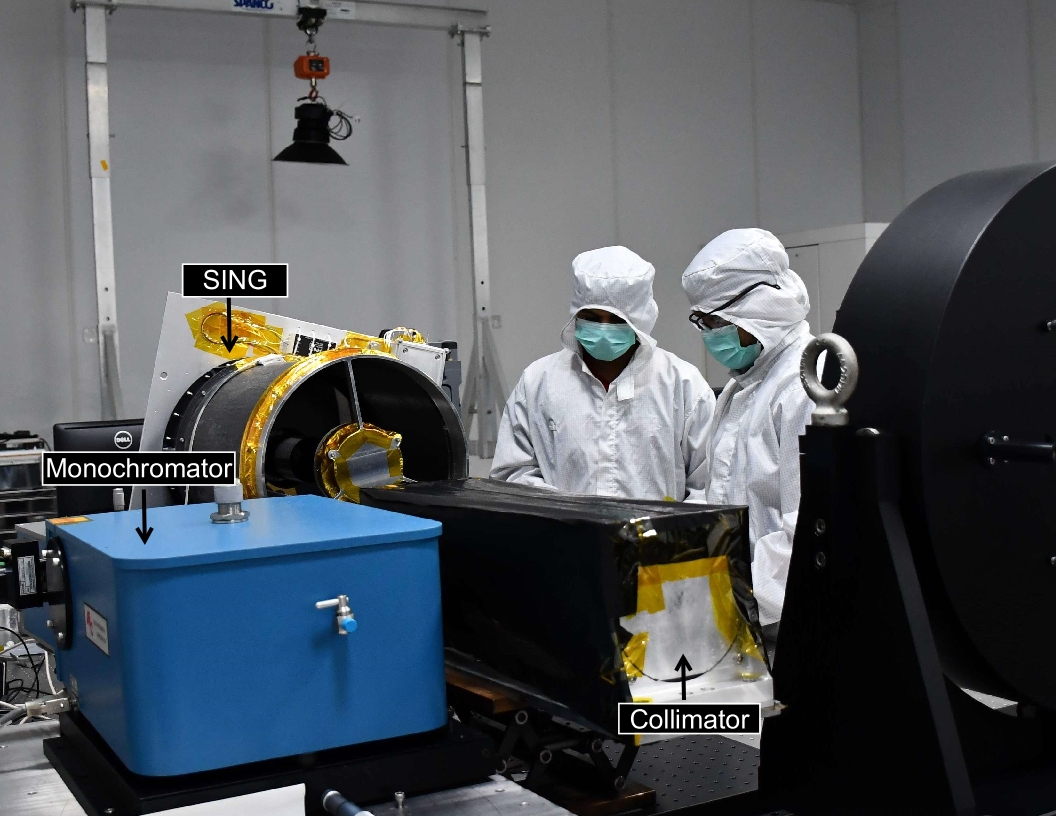}
    \vspace{1.5mm}
    \caption{Test setup for calibration and resolution measurement.}
    \label{fig:wav_setup}
\end{figure}
\end{minipage}
\hfill
\begin{minipage}[t]{0.49\textwidth}
\begin{figure}[H]
    \centering
\includegraphics[width=\linewidth]{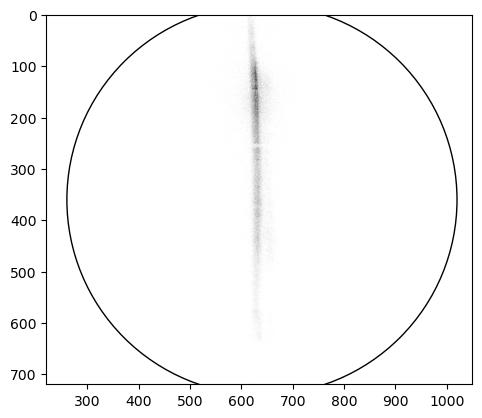}
\caption{2200\AA\ line with 15\AA\ bandwidth imaged in photon-counting mode.}
    \label{fig:line}
\end{figure}
\end{minipage}

The spectral resolution and dispersion measurement were performed using the monochromator collimator setup (Fig.~\ref{fig:wav_setup}). To perform this measurement, we used an Acton VM-504 monochromator with a deuterium source, which has a dispersion value of 2.1 nm/mm at the exit slit. We set the monochromator exit slit width to a wavelength bandwidth of 15 \AA\ ($\sim$0.7 mm slit width). To verify the bandwidth, we used MAYA UV spectrograph\footnote{https://www.oceaninsight.com/products/spectrometers/high-sensitivity/Custom-Configured-maya2000-pro--series/} to measure the FWHM of the light from the collimator output, and found it to be 15 \AA. 

We have recorded the spectra of light coming from the collimator using SING in the photon-counting mode at different wavelength settings in the monochromator, starting from 2700 \AA\ till 2000 \AA\ with an interval of 100 \AA. Each spectrum was recorded for a duration of 2 minutes, and the centroid values generated by the detector were collected through the RS485 interface and stored on a computer for analysis. We also recorded the spectra from the collimator using the MAYA UV spectrograph as a reference.

Using the centroids values, we generated the 2D images for each wavelength (Fig.~\ref{fig:line} shows 2200 \AA\ line 2D image recreated from the photon counting x-y position data).
The 2D image was then converted to 1D spectra for 50-pixel vertical bins to determine the peak for each wavelength and the pixel position of each wavelength across the fields. After which, we deconvolved the 15 \AA\ input from the 1D spectra of each wavelength to determine the resolution as a function of wavelength at different positions on the detector. Fig.~\ref{fig:resolution_plot} shows the variation of resolution as a function of wavelength. We also recorded the pixel position of each wavelength recorded on the detector to get the wavelength calibration information (Fig.~\ref{fig:pos_cal}).

\begin{minipage}[t]{0.49\textwidth}
    \begin{figure}[H]
    \centering
    \includegraphics[width=\linewidth]{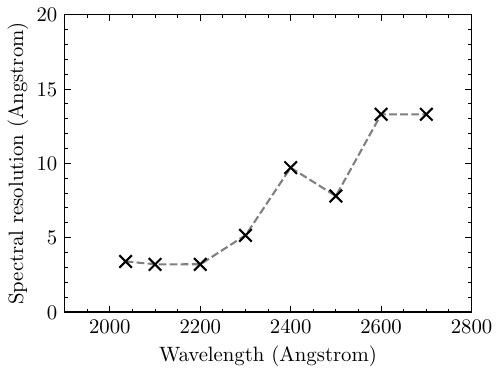}
    \caption{Resolution as a function of wavelength for the central field.}
    \label{fig:resolution_plot}
    \end{figure}
\end{minipage}
\hfill
\begin{minipage}[t]{0.49\textwidth}
    \begin{figure}[H]
    \centering
    \includegraphics[width=\linewidth]{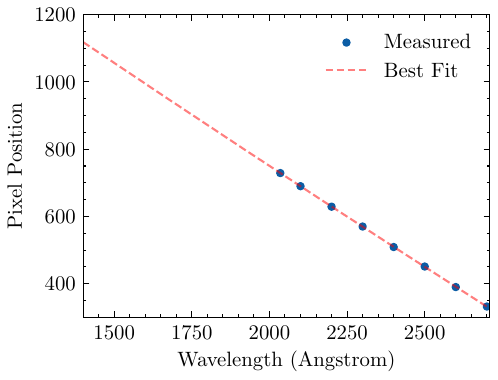}
    \caption{Wavelength vs pixel position measurement.}
    \label{fig:pos_cal}
\end{figure}
\end{minipage}

\section{Observation strategy}

The SING is planned to operate from the experimental module of the space station. SING will be fixed onto the payload adapter of the experimental module of the space station and will not have the freedom for pointed observations, so SING will operate in a scanning mode. As the CSS orbits the Earth, it will scan the regions of the sky constrained by the orbital parameters of the space station. As the platform revolves around the Earth, SING will generate spectra of different regions of the sky. Each orbit becomes a canvas of systematic observation in the wavelength range $1400-2700$ \AA. The information regarding the pointing during observation will be provided by the star camera aligned to the SING. In addition to this, SING will also get the pointing information from CSS.

 In the scanning mode of observation, each detected photon event will be tagged with x-y detector position, time and pointing data from the star camera and the observation is planned only during orbital nights. So later, at the ground, the spectra of any region that SING has observed can be generated by integrating the photon event record of that region captured over multiple scanning passes. Data generated per orbit is a maximum of 45 MB for a 90-min LEO orbit; these include the x-y centroids data, time stamps, housekeeping data, and attitude information with the required headers and footers, which will be transmitted to the ground station through the satellite/platform bus. 
\begin{figure}[H]
    \centering
    \includegraphics[width=\textwidth]{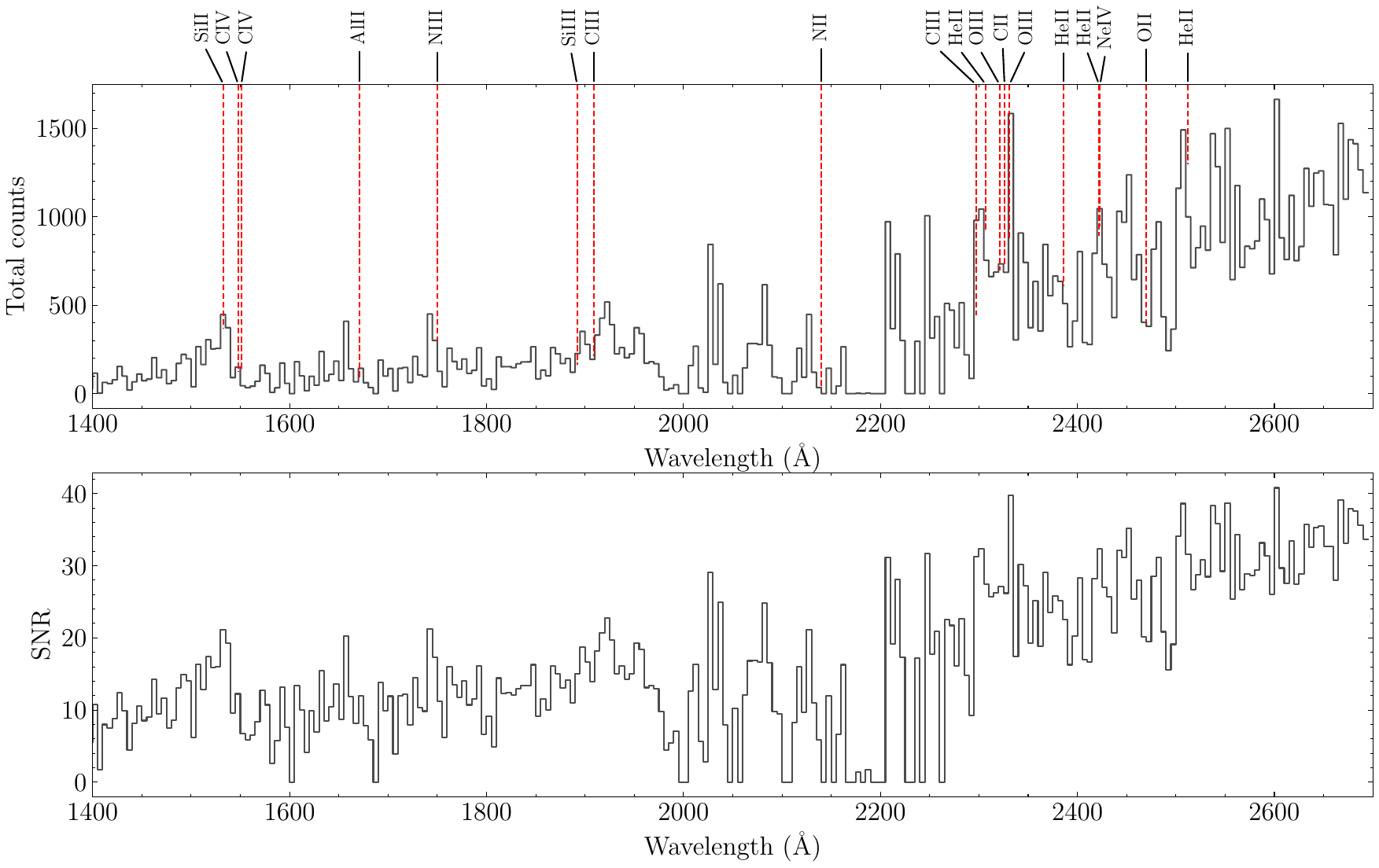}
\caption{Simulation of Crab Nebula spectrum and the associated SNR obtained by SING for 100-sec exposure.}
    \label{fig:crab_plot}
\end{figure}

\begin{figure}[h]
\centering
\includegraphics[width=0.7\textwidth]{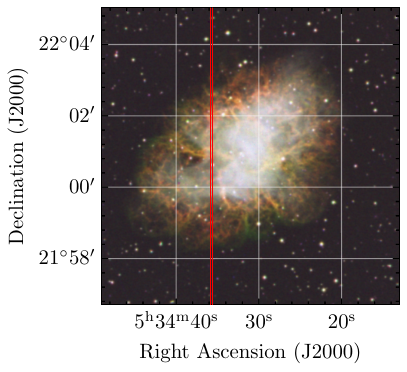}
\caption{Image of the Crab Nebula with slit field marked in red for reference.}
\label{crab}
\end{figure}

The mission is planned for $12-24$ months of observations, during which SING will create a spectral map of the NUV sky, constrained by the orbital parameters. Fig.~\ref{fig:crab_plot} shows the simulated spectrum and the SNR for the Crab Nebula (Fig.~\ref{crab} shows the Crab Nebula field) over 100-sec exposure,  which can be achieved after 100 passes in a scanning mode of observation over the entire 4 arc minute field\footnote{Here we have assumed that the slit is oriented perpendicular to the direction of motion, and the scan was done at a slew rate of 4 arc minute per sec.}; the CIV line in the spectra has an SNR $>$ 12. The data input for simulation was obtained from the IUE \cite{iue_nasa} observations of the Crab Nebula. This data was binned to a resolution of 5 \AA\ of SING, and using the effective area of SING and exposure time, the photon count was calculated for each resolution element.  
 In the simulation, we did not include the dark counts and backgrounds. The UV backgrounds, such as zodiacal light and airglow, depend on parameters such as the date and time of observation. Zodiacal light will depend on the angle from the Sun and ecliptic latitude of observation, while airglow depends on the local time of observation\cite{Leinert,2010ApJ...724.1389M}. Also, the noise for an MCP-based photon counting detector is contributed mainly by the photon noise, dark counts and cosmic ray interactions\cite{2018Ap&SS.363...63C}. The dark counts and cosmic ray interactions can be characterized in orbit and taken care of during data reduction.

The power required for operating SING is under 15 W and will be provided by the space platform through the unregulated 28 V DC line. The operating temperature will be maintained by means of MLI and heaters. The current interface we have implemented is RS485 for instrument calibration and testing. The RS485 module is connected directly to the photon-counting detector. We are currently implementing the MIL 1553B communication using a custom STM32-based onboard controller (OBC) and Holt 1553B transceiver module. Once the electronics interface test is completed, we will be integrating the OBC and 1553B communication modules into the SING electronics enclosure. The additional required attitude information including the time will be provided by the CSS through the 1553 BUS.

\section{Conclusions and future work}

\begin{figure}[H]
\centering

  \begin{tabular}{@{}cc@{}}
    \begin{subfigure}[b]{0.49\textwidth}\includegraphics[width=\textwidth]
    {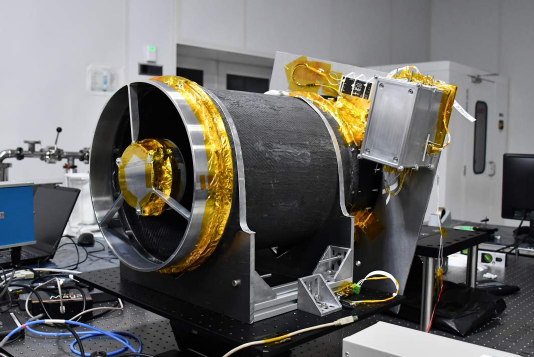} \end{subfigure} &
    \begin{subfigure}[b]{0.49\textwidth}\includegraphics[width=\textwidth]{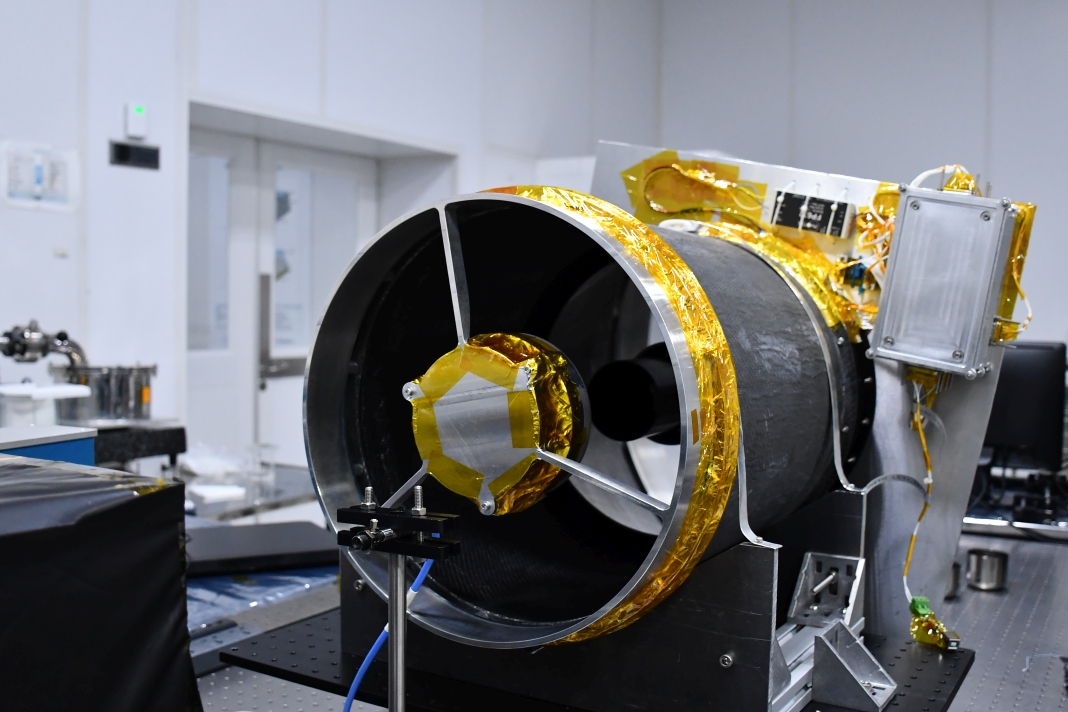}\end{subfigure}   
\end{tabular}
\caption{SING after assembly in class 100,000 room at the MGKM Lab.}
  \label{fig:sing_image}
\end{figure}

The assembly, alignment, and calibration of the spectrograph for the spectroscopic investigation of nebular gas (SING) have been completed. Fig.~\ref{fig:sing_image} shows the assembled SING instrument on the optical table in the MGKM Lab. The design was very cost-effective, and we were able to manufacture most of the components in-house at IIA facilities. The design was done to meet the flight requirements of the  Long March-2F manned spaceship launch loads. The instrument is able to achieve the best spectral resolution of $3.7$ \AA\, and a spatial resolution of 1.33 arc minute. We performed the wavelength calibration from 2000 \AA\ to 2700 \AA\ in the air. The complete test in the vacuum chamber will be performed in the near future. The current electronics interface implemented in SING is RS485. We are developing the 1553 MIL interface for communication with the CSS station, however this interface can easily be modified to meet the requirement of any stable platform. To ensure that the payload is able to withstand all launch loads and operate in the space environment, the final flight qualification vibration and thermal-vacuum tests \cite{StarSensor} will be performed in the MGKM Lab, following the Long March-2F manned spaceship flight requirements. Further mission planning and simulations are under development; based on this, the mounting point on the CSS will be determined. We are currently working on orbital simulation, exposure time calculator, sky coverage analysis, and data reduction pipeline for the SING mission.

\section*{Acknowledgements}

We specially thank Mr. Burman Rahuldeb, Mr. Ismail Jabillulah and Mr. G. Nataraj who manufactured the mirrors for SING at the IIA photonics lab. The authors thank Mr. S.~Kathiravan for helping us complete the project at the MGKM Lab, CREST Campus, IIA, Bangalore. We express our sincere gratitude to Mr. Ajin Prakash of Arksa Research Labs for his invaluable suggestions and assistance. We also thank Mr. Prasobh P. and Mr. Francis from IIA mechanical workshop, and the technicians Mr. Vishnu and Mr. Suresh from MGKM Lab. MS acknowledges the financial support by the DST, Government of India, under the Women Scientist Scheme (PH) project reference number SR/WOS-A/PM-17/2019. Some of the data presented in this paper were obtained from the Mikulski Archive for Space Telescopes (MAST). Space Telescope Science Institute (STScI) is operated by the Association of Universities for Research in Astronomy, Inc., under National Aeronautics and Space Administration (NASA) contract NA S5-26555. Support for MAST for non-Hubble Space Telescope (HST) data is provided by the NASA Office of Space Science via grant NNX09AF08G and by other grants and contracts.

\begin{appendices}
\section{Code, data, and materials availability} 
Data from the experiment will be available upon request.

\section{Funding declaration} 

This research was made possible through financial support from the Indian Institute of Astrophysics (IIA) under the Department of Science and Technology (DST) in India. 

\section{Author contribution declaration}

BCP, BK, MS\textsuperscript{1}, and JM authored the primary manuscript text. BCP and BK were responsible for the system design and testing. MB and SJ\textsuperscript{2} were responsible for the assembly of the electronics. Testing of the detector was a collaborative effort involving BCP, BK, and SJ\textsuperscript{1}. The project received essential oversight and funding support from JM, RM, and MS\textsuperscript{2}, who served as project supervisors.
\section{Declaration of competing interest} 

The authors declare that they have no known competing financial interests or personal relationships that could have appeared to influence the work reported in this paper.

\end{appendices}


\bibliography{report}   

\end{document}